\PassOptionsToPackage{table,dvipsnames}{xcolor}

\documentclass[sigconf, screen=true]{acmart}

\acmSubmissionID{2146}

\usepackage{algorithm}
\usepackage{algpseudocode}

\usepackage[inline]{enumitem}
\usepackage{acronym}
\usepackage[skip=2pt]{caption}
\usepackage{booktabs}
\usepackage{multirow}
\usepackage{graphicx}
\usepackage{tabularx}
\usepackage{colortbl}

\usepackage{makecell}
\usepackage{titlesec}
\usepackage[colorinlistoftodos]{todonotes}
\usepackage{array}
\usepackage{dsfont}

\usetikzlibrary{arrows.meta}
\usetikzlibrary{shapes.geometric}
\usetikzlibrary{shapes.symbols}

\AtBeginDocument{%
  }

\titleformat{\subsubsection}
  {\normalfont\large\bfseries}{\thesubsubsection}{1em}{}

\author{Mohanna Hoveyda}
\orcid{0009-0003-8027-6575}
\affiliation{%
  \institution{Radboud University}
  \city{Nijmegen}
  \country{The Netherlands}
}
\email{mohanna.hoveyda@ru.nl}

\author{Harrie Oosterhuis}
\orcid{0000-0002-0458-9233}
\affiliation{%
  \institution{Radboud University}
  \city{Nijmegen}
  \country{The Netherlands}
}
\email{harrie.oosterhuis@ru.nl}

\author{Arjen P. de Vries}
\orcid{0000-0002-2888-4202}
\affiliation{%
  \institution{Radboud University}
  \city{Nijmegen}
  \country{The Netherlands}
}
\email{arjen.devries@ru.nl}

\author{Maarten de Rijke}
\orcid{0000-0002-1086-0202}
\affiliation{%
  \institution{University of Amsterdam}
  \city{Amsterdam}
  \country{The Netherlands}
}
\email{m.derijke@uva.nl}

\author{Faegheh Hasibi}
\orcid{0009-0006-9986-482X}
\affiliation{%
  \institution{Radboud University}
  \city{Nijmegen}
  \country{The Netherlands}
}
\email{faegheh.hasibi@ru.nl}

\copyrightyear{2025}
\acmYear{2025}
\setcopyright{cc}
\acmConference[SIGIR '25]{Proceedings of the 48th International ACM SIGIR
Conference on Research and Development in Information Retrieval}{July 13--18,2025}{Padua, Italy}
\acmBooktitle{Proceedings of the 48th International ACM SIGIR Conference on
Research and Development in Information Retrieval (SIGIR '25), July 13--18,
2025, Padua, Italy}
\acmDOI{10.1145/3726302.3730351}
\acmISBN{979-8-4007-1592-1/2025/07}

\makeatletter
\gdef\@copyrightpermission{
\begin{minipage}{0.3\columnwidth}
\href{https://creativecommons.org/licenses/by/4.0/}{\includegraphics[width=0.90\textwidth]{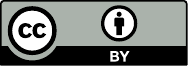}}
\end{minipage}\hfill
\begin{minipage}{0.7\columnwidth}
\href{https://creativecommons.org/licenses/by/4.0/}{This work is licensed under a Creative Commons
Attribution International 4.0 License.}
\end{minipage}
\vspace{5pt}
}
\makeatother

\ccsdesc[500]{Information systems~Information retrieval; Search engine architectures and scalability}

\keywords{Adaptive Information Systems,
Graph-Based Orchestration}

\begin{document}

\title{Adaptive Orchestration of Modular~Generative~Information~Access~Systems}

\begin{abstract}
Advancements in large language models (LLMs) have driven the emergence of complex new systems to provide access to information, that we will collectively refer to as modular generative information access (GenIA) systems.
They integrate a broad and evolving range of specialized components, including LLMs, retrieval models, and a heterogeneous set of sources and tools. 
While modularity offers flexibility, it also raises critical challenges: 
How can we systematically characterize the space of possible modules and their interactions?
How can we automate and optimize interactions among these heterogeneous components? And, how do we enable this modular system to dynamically adapt to varying user query requirements and evolving module capabilities?

In this perspective paper, we argue that the architecture of future modular generative information access systems will not just assemble powerful components, but enable a self-organizing system through real-time adaptive orchestration -- where components' interactions are dynamically configured for each user input, maximizing information relevance while minimizing computational overhead. We give provisional answers to the questions raised above with a roadmap that depicts the key principles and methods for designing such an adaptive modular system. We identify pressing challenges, and propose avenues for addressing them in the years ahead.
This perspective urges the IR community to rethink modular system designs for developing adaptive, self-optimizing, and future-ready architectures that evolve alongside their rapidly advancing underlying technologies.

\end{abstract}

\maketitle

\section{Introduction}
Search engines have been the dominant gateway to the vast amount of information online for decades by now. However, traditional search systems are increasingly inadequate as the user's information demands evolve \cite{allan2024future, DBLP:conf/sigir/Zhai24, 10.1145/3626772.3661375}. Today’s users no longer settle for a static list of result documents to be manually sifted through; they expect direct answers to questions that might require complex reasoning~\cite{DBLP:conf/eacl/WellerLD24, DBLP:conf/acl/MalaviyaSCLT23}, a synthesis of different pieces or modalities of information~\cite{zhou-etal-2022-answer, DBLP:conf/emnlp/WuJ0RPWN24}, or even carrying out tasks or transactions on behalf of them -- all delivered through a seamless interactive multimodal interface~\cite{allan2024future}.

Recent breakthroughs in the development of large language models (LLMs) have provided highly promising advancements in multi-turn dialogue capabilities, improved intent detection, and richer content synthesis. 
These advancements open new opportunities for developing future information retrieval (IR) systems and conversational assistants, and can help tackle some of the persistent challenges in information access in the era of generative AI~\cite{DBLP:conf/sigir/Zhai24, allan2024future}. We will refer to these future systems as \emph{Generative Information Access (GenIA) systems}, to emphasize our focus on the long-standing problem of information access while acknowledging the importance of generative AI to achieve this objective.

While future GenIA systems could be monolithic LLMs, we see many advantages for a new system architecture, consisting 
of ensembles of diverse components, with clearly identifiable roles, that interact with each other in order to generate answers to user queries~\citep{compound-ai-blog, DBLP:conf/nips/LewisPPPKGKLYR020, DBLP:conf/iclr/LiZCDJPB24, DBLP:conf/nips/LuPCGCWZG23, DBLP:journals/corr/abs-2305-17066}.\footnote{We purposely write \emph{identifiable} instead of \emph{identified}, since our perspective posits that the roles of components should be learned and adaptive.}
E.g., retrieval can employ a variety of models, including sparse, dense, and graph-based models, each tailored for particular tasks or corpus types~\cite{DBLP:journals/ftir/RobertsonZ09, DBLP:conf/emnlp/KarpukhinOMLWEC20,DBLP:conf/sigir/LiAH24, DBLP:journals/tkde/PanLWCWW24}. Post-retrieval modules further refine the information being passed to the language model, leveraging natural language inference filters~\cite{DBLP:conf/iclr/YoranWRB24}, atomic-fact extractors, or even controlled noise injection~\cite{cuconasu2024power} to ensure more robust answer formulation.
For complex queries, such as those requiring multi-hop reasoning~\cite{DBLP:conf/emnlp/Yang0ZBCSM18} or logical inference \cite{DBLP:conf/eacl/WellerLD24}, approaches such as neuro-symbolic pipelines expand the space of possible modules with symbolic solvers~\cite{Olausson_2023, ganguly2024proof}. When queries demand task execution, a variety of tools can be incorporated~\cite{yuan2024craft}. Likewise, domain-specific tasks may be addressed using specialized or variably scaled LLMs \cite{yin-etal-2024-enhancing, barman2025largephysicsmodelscollaborative}.

Taken together, these components and implementations highlight a shift towards ensembles of specialized, interoperable modules that collectively enable effective information access or conversational assistance. A variety of frameworks~\cite{chase2022langchain, DBLP:conf/icml/ZhugeWKFKS24, DBLP:conf/nips/LiHIKG23, DBLP:journals/corr/abs-2405-13576, DBLP:journals/corr/abs-2407-21059, DBLP:conf/sigir/LianHC0WWJFLWC23} has been proposed to chain specialized modules into end-to-end systems. However, these solutions often require manually designed pipelines \cite{chase2022langchain, DBLP:journals/corr/abs-2405-13576} or rely on a static \emph{orchestration} of modules \cite{DBLP:conf/icml/ZhugeWKFKS24}.

In this perspective paper, we posit that modularity in GenIA systems introduces a highly important novel research direction into our field:
the development of dynamic system designs for information retrieval systems that automatically adapt the modules in its pipeline and the interaction between them.
Focusing research on learning the optimal interactions between modules is crucial to advancing modern, generative information access systems.
We propose that an adaptive orchestration framework should dynamically coordinate module selection and sequencing based on the properties of the query, the capabilities of the modules, and computational constraints. 
Thereby, such frameworks will not merely improve response quality but balance quality and accuracy with latency and other costs.
Orchestration frameworks should learn from real-time feedback -- i.e., through user satisfaction scores, error detection, or runtime metrics -- such that the orchestration strategy can be continuously refined, over time. 
Crucially, this adaptability also paves the way for incremental integration of new modules without necessitating a complete overhaul of existing system infrastructure, a highly desirable property from the view of operations.
Central to this vision are three key questions:
\begin{enumerate}[leftmargin=*]
\item
\emph{How can we rigorously define the space of possible modules?}
\item
\emph{How can we model, automate, and optimize interactions among these heterogeneous components?}
\item
\emph{How should this modular system dynamically adapt to varying user query requirements and evolving module capabilities?}
\end{enumerate}

\noindent%
We offer provisionary answers to these questions through the proposal of a novel framework that addresses these challenges.
Our framework relies on a graph-based representation~\cite{DBLP:conf/icml/ZhugeWKFKS24}, where nodes and edges correspond to modules that represent the subtasks of a pipeline, what agents/tools perform these tasks and the usage of information resources.
 To achieve adaptive orchestration of systems planning and reinforcement learning algorithms can be used to find and construct the most effective pipeline-design graph for each incoming user request.
Thereby, a balance between effectiveness and efficiency is optimized.

After discussing related work in Section~\ref{sec:relatedwork}, our perspective on the future of information retrieval systems is detailed in Section~\ref{sec:perspective}.
We propose a general framework by which our perspective can be realized in Section~\ref{sec:framework}, which also lays out different implementation choices.
In Section~\ref{sec:Instantiation} we provide an example instantiation of our proposed framework for an adaptive question answering system that uses a Contextual Multi-Armed Bandit (CMAB) for optimization~\citep{li2010contextual}.
Section~\ref{sec:results} provides experimental results that evaluate this instantiation and reveal it successfully adapts to balance efficiency and effectiveness.
Finally, we discuss our perspective and proposed approach in the concluding Section~\ref{sec:conclusion}.

\section{Background}
\label{sec:relatedwork}
\subsection{Multi-agentic LLM orchestration}
Several studies have proposed frameworks for enabling multiple language model-based agents and related modules to communicate to solve tasks \citep{DBLP:conf/iclr/HongZCZCWZWYLZR24, DBLP:journals/corr/abs-2308-08155, DBLP:journals/tmlr/AdolphsBBHCEHKRSS22}.
Most work focuses on orchestration without further optimization of the structure of these agents \citep{DBLP:journals/corr/abs-2308-08155, DBLP:journals/corr/abs-2305-17066}.
Inspired by Minsky's \textit{society of minds} (SoM) \citep{minsky1988society}, which describes how smaller parts of a system can collaborate to achieve a goal, \citet{DBLP:journals/corr/abs-2305-17066} suggest a shift from relying on optimizing a single model for solving a task to the optimization of information flow between two or more models. %
\citet{DBLP:conf/icml/ZhugeWKFKS24} propose GPTSwarm to optimize a society of language model-based agents. GPTSwarm structures a system made of multiple agents as a graph, and every computational operation (e.g., querying a large language model with a prompt) is represented as a node within that graph.
The dual-level optimization approach of GPTSwarm optimizes prompts at the node level while also enhancing the flow of information by pruning out edges and nodes that are not found useful. While innovative, GPTSwarm proposes an optimization framework that only yields a single static final graph which fails to adapt to different user needs and queries and the varied capabilities of its underlying components. 
\vspace{-1mm}
\subsection{Mixture of experts (MoE)}
A paradigm that is related to, but different from the society of mind (SoM) framework is the \textit{mixture of experts (MoE)} approach \cite{DBLP:conf/iclr/LepikhinLXCFHKS21, DBLP:conf/icnn/SzymanskiL93}.
MoE architectures typically partition model capacity across multiple \textit{expert} networks, each specialized in handling specific parts of the input space. A gating or routing network then determines which expert(s) to activate for a given input, facilitating more efficient model scaling and potentially better task performance. Recent advances in MoE have focused on improving routing strategies to reduce computational overhead, balance the load among experts, and refine expert specialization \cite{zhou2022mixture,DBLP:journals/corr/abs-2401-04088}.

While both use modular components and routing, MoE learns expert specialization implicitly through training and uses internal vector-based communication, whereas in SoM, modules are explicitly designed for predefined functions with natural language communication enabling higher interpretability.  Additionally, SoM allows for heterogeneous modules with diverse capabilities, unlike MoE's typically homogeneous expert structure.

\vspace{-1mm}
\subsection{Adaptivity in IR systems}
The need for adaptivity in IR systems has received increasing attention in recent research, with varying interpretations and objectives across different contexts.
\citet{DBLP:conf/sigir/ParryGC24} introduce the notion of adaptively selecting the number of in-context-learning (ICL) examples for queries with varying levels of complexity. \citet{DBLP:conf/sigir/DengLZYC24} advocate for a pro-active conversational agent that can adaptively decide when to intervene and engage the user. In the domain of retrieval-augmented generation, when to retrieve and when to rely on an LLM's parametric knowledge based on a given query is a challenging and interesting problem \cite{su-etal-2024-dragin, jiang-etal-2023-active, DBLP:conf/acl/MallenAZDKH23}. \citet{DBLP:conf/acl/MallenAZDKH23} propose a binary threshold-based framework that distinguishes whether questions contain popular or long-tail entities, and decides whether to retrieve accordingly. Following this work, several other approaches have been proposed to tackle this problem using an LLM as a classifier \cite{DBLP:conf/naacl/JeongBCHP24} or guided by a reinforcement learning-based reward model \cite{DBLP:conf/iclr/AsaiWWSH24, DBLP:conf/coling/TangGLDLX25}.

\section{Perspective: Designs of Future GenIA Systems Should be Dynamic, Modular and Adaptive}
\label{sec:perspective}

For a long time, information retrieval systems have relied on static architectures: monolithic search engines, predefined pipelines, and rigid workflows that process user queries in a pre-determined manner. While recent advancements in LLM technologies have enabled the integration of diverse models and tools within these pipelines, the underlying interaction between these components remains largely inflexible, lacking adaptivity and failing to dynamically adjust based on the specific requirements of each input.
As the field continues to advance with the introduction of novel and sophisticated models and tools, we believe the time has come for a new architecture of information access systems, where different modules communicate with each other interactively to construct customized information system pipelines, while adapting their interactions based on user input for high throughput, efficiency, and effectiveness.

The need for such adaptability is already evident. User inputs vary widely in their complexity -- some require simple fact retrieval, while others demand logical reasoning, disambiguation, or multi-step synthesis. 
Certain queries may involve task execution, such as booking a flight, while others demand specialized capabilities, like generating code or synthesizing insights from multiple research papers. Yet, most current systems treat all inputs uniformly, applying a static pipeline using the same computational resources, failing to differentiate between trivial and intricate queries. This one-size-fits-all approach is inefficient and suboptimal, leading to unnecessary computational overhead for simple tasks and inadequate depth or poorly suited strategies for complex ones.

What we need is a new architecture for information access systems that does not simply retrieve information but orchestrates an evolving ensemble of specialized models and tools, each playing its role based on its relevance and utility in the moment. A user submits a query, and the system determines which modules to invoke, how they should communicate, and how to optimize their collective output; similar to assembling a team of experts on the fly.

Achieving this vision introduces several challenges.
First, we need a framework that allows modules to interact dynamically rather than through fixed orchestration rules. This requires a shift towards  expressive \emph{graph-based interaction models}, where modules form transient communication patterns tailored to each query.
Second, \emph{adaptivity} must be built into the system itself, such that it learns when to engage certain components, how to weight their contributions, and when to optimize for speed versus depth of reasoning. This demands an optimization mechanism capable of adjustment in real-time, ensuring that computation is allocated efficiently and based on the query context.
\noindent
Third, as modular systems continue to evolve, new components will emerge, and existing ones will improve. A truly future-ready Information Access system must not only adapt to individual queries but also accommodate continuous changes in its own architecture. This requires mechanisms for dynamically integrating new modules, updating orchestration strategies, and ensuring that the system remains robust as its components evolve.
Without such adaptability, the benefits of modularity remain an unrealized promise, a collection of powerful components constrained by a rigid system structure.

The architecture of a GenIA system should go beyond merely assembling a few sophisticated modules; instead, it needs a self-optimizing, adaptive orchestration framework that allows various types of computational processes (next token prediction, retrieval, classification, symbolic reasoning, etc.) to interact meaningfully, in real-time.

In the next section, we propose our framework, to pave the way for the realization of this vision.

\begin{table*}[t]
\small
\centering
\setlength{\tabcolsep}{1.2mm}
\caption{Overview of the different module types and their subcategories in our framework.}
\label{tab:moduleoverview}
\footnotesize
\begin{tabularx}{\textwidth}{l l X l }
\toprule
\bf Type & \bf Subcategory & \bf Description & \bf Visualization
\\
\midrule
\multirow{2}{*}{Tasks}
&  \raisebox{1.5mm}{Standalone} &
\raisebox{1.5mm}{Simple operations that do not require interactions between multiple agents and tools.}
& \scalebox{0.85}{  \tikz \node[fill=BlueGreen, text=white, signal, draw=BlueGreen, minimum height=0.5cm, minimum width=1cm] {\footnotesize Standalone task};}  

\\
\cmidrule{2-4}
& \raisebox{1.5mm}{Complex} & \raisebox{1.5mm}{Multi-step operations requiring agents or interactions between multiple modules.} %
& 
\scalebox{0.85}{\tikz \node[fill=RoyalBlue, text=white, signal, draw=RoyalBlue, minimum height=0.5cm, minimum width=1cm] {\footnotesize Complex task}; } 

\\
\midrule
\multirow{2}{*}{Executor}
& \raisebox{1.5mm}{Agents} & \raisebox{1.5mm}{General execution units (i.e., foundation models) that are widely applicable and can use tools.}  & 
\scalebox{0.85}{\tikz \node[fill=Maroon, text=white, rectangle, draw=Maroon, minimum height=0.5cm, minimum width=1cm] {\footnotesize Agent};  }
\\
\cmidrule{2-4}

&  \raisebox{1.5mm}{Tools} & \raisebox{1.5mm}{Specialized execution units designed for specific tasks with limited applicability.}
& 
\scalebox{0.85}{\tikz \node[fill=Bittersweet, text=white, rectangle, draw=Bittersweet, minimum height=0.5cm, minimum width=1cm] {\footnotesize Tool};  }

\\
\midrule
\raisebox{1.5mm}{Resources}
& \raisebox{1.5mm}{(Properties)} & \raisebox{1.5mm}{Sources of data/information; categorized on the following properties: \emph{Structure}, \emph{Modality} and \emph{Accessibility}.} & 
\scalebox{0.85}{\tikz \node[fill=OliveGreen, text=white, ellipse, draw=OliveGreen, minimum height=0.5cm, minimum width=1cm] {\footnotesize Resource};  }
\\
\bottomrule
\end{tabularx}

\end{table*}

\section{Adaptive Module Orchestration Framework}
\label{sec:framework}

This section introduces our proposed 
architecture for modular \textit{GenIA} systems that dynamically adapts to user queries.
Section~\ref{sec:framework:design} describes our main design principles and goals.
In Section~\ref{sec:framework:modules} we propose a categorization of modules that make up system designs.
Lastly, Section~\ref{sec:framework:pipeline} discusses how system designs can be dynamically constructed to adapt to individual user queries.

\subsection{System design principles and goals}
\label{sec:framework:design}

The main goal of our framework is to produce an information retrieval system that optimizes a typical balance of effectiveness and efficiency.
In other words, we aim to provide a system that is effective at providing high quality responses to user queries, while not incurring too much cost (typically in terms of response time or computational costs).
Accordingly, the design of our system should be chosen to optimize our goal, which gives rise to three key questions:
\begin{itemize}[leftmargin=*]
    \item \textit{What tasks should be executed and in what order?}
    \item \textit{What tools and agents should be used to execute each task?}
    \item \textit{What sources of information should be consulted?}
\end{itemize}
Together, the answers to these questions describe the inner pipeline of a system, and thus, they capture its behavior and cost to operate.

Today's paradigm to information retrieval system design is to answer these questions once and create a static pipeline that performs the same procedure for every user query.
Conversely, we propose a radically different approach and argue that the answers to these design questions should be reconsidered for every user query.
Thereby, the system's design is re-constructed on-the-fly and can adapt to pursue the best effectiveness-efficiency trade-off for individual user queries.

Our dynamic approach means that systems are no longer designed as rigid pipelines, instead, the task of the practitioner is now to develop a space of possible designs that can be searched, with designs that can be quickly constructed and executed.
Our framework enables this through a modular approach, that describes the building blocks that can comprise pipelines; and an adaptive procedure to find, learn and construct pipelines in response to queries.

\subsection{Modules: Tasks, tools, agents and resources}
\label{sec:framework:modules}

Since our framework is designed for systems of the future, we aim to encapsulate all existing systems and all of their popular extensions.
In order to be future-proof, we apply generic concepts and categories that should remain relevant to future advancements.
This also applies to our modular approach, as it only defines on three types modules, each with their own subcategories.
Our module types are conceptualized to match the three key design questions posed in Section~\ref{sec:framework:design}:
\begin{enumerate*}[label=(\roman*)]
    \item \emph{tasks} (what is done),
    \item \emph{executors} (what does it), and
    \item \emph{resources} (where information comes from).
\end{enumerate*}
Table~\ref{tab:moduleoverview} provides an overview of our module categories; the remainder of this section describes each category separately.
Table~\ref{tab:node_space} provides a list of examples per module type.

\subsubsection{Task modules}
The first type of modules: tasks ($\mathcal{T}$), represent the procedures that can be performed inside a system pipeline.
These modules thus describe \emph{what the pipeline does}, but do  \emph{not} indicate what performs the procedures or what resources are used.
Instead, tasks can have requirements, i.e., which types of executors can perform them, or what resources are needed to complete them.
We propose two subcategories for this module type: \emph{standalone} and \emph{complex} tasks.

\textbf{Standalone tasks} refer to simple operations that can be executed independently without requiring interactions between multiple executors or complex agents.
In general, these are tasks for which specialized tools exist that can only perform that single task.
Examples include but are not limited to tasks of \emph{retrieval}, \emph{intent classification}, \emph{query rewriting}, and \emph{direct generation}.
Due to their self-contained nature, standalone tasks can be efficiently executed with minimal dependencies, making them well-suited for constructing system designs that are lightweight and fast.

\textbf{Complex tasks} refer to more complex operations that can only be executed by multiple executors or complex agents or procedures.
In contrast with standalone tasks, these tasks generally involve multiple interdependent operations that cannot be performed by specialized tools.
This often means a single action is insufficient to produce a meaningful result, necessitating multi-step reasoning or iterative refinement.
Examples include \emph{Retrieval Augmented Generation (RAG)}~\cite{DBLP:conf/nips/LewisPPPKGKLYR020}, \emph{Interleaving Retrieval with Chain-of-Thought Reasoning (IRCOT)}~\cite{DBLP:conf/acl/TrivediBKS23} , and \emph{Logical Inference via Neurosymbolic Computation (LINC)}~\cite{Olausson_2023}.
In general, complex tasks are less efficient but more effective, and thus, more appropriate when standalone tasks are ineffective.

{
\newcolumntype{P}[1]{>{\raggedright\sloppy\arraybackslash}p{#1}}
\begin{table*}[h]
    \centering
    \small 
    \setlength{\tabcolsep}{2pt} 
    \caption{Non-exhaustive overview of potential choices for the modules per type (possible nodes in the graphs).}
    \label{tab:node_space}
    \setlength{\aboverulesep}{0pt}
    \setlength{\belowrulesep}{0pt}
    \setlength{\extrarowheight}{.75ex}
    \footnotesize
    \begin{tabularx}{\textwidth}{@{}P{0.10\textwidth} P{0.18\textwidth} X@{}}
        \toprule
        \textbf{Node type} & \textbf{} & \textbf{Description} \\
        \midrule
        \multirow{18}{*}{\textbf{Tasks}} 
        & \multicolumn{2}{l}{\cellcolor{gray!15} \textbf{\textit{Standalone}}} \\ \cmidrule{2-3}
        & \textbf{Generation} & Generate text (or image and other modalities) given a certain input \cite{DBLP:journals/corr/abs-2302-13971, brown2020language, DBLP:conf/iclr/DarcetOMB24} \\
        & \textbf{Retrieval} & Retrieve pieces of information given a certain input \cite{DBLP:journals/ftir/RobertsonZ09, DBLP:conf/emnlp/KarpukhinOMLWEC20,DBLP:conf/sigir/LiAH24} \\
        & \textbf{Query rewriting} & Reformulate/expand queries to improve retrieval performance \cite{10.1145/3543873.3587678, zhou-etal-2023-unified} \\
        & \textbf{Intent recognition} & Categorize user input into predefined intents or discover novel intents \cite{zhou-etal-2023-probabilistic, arora-etal-2024-intent} \\
        & \textbf{Asking questions} & Asking clarifying questions proactively to engage the user to help the retrieval process \cite{aliannejadi-etal-2021-building, DBLP:conf/www/0002SARL24} \\
        & \textbf{Entity Linking} & Identifying entity mentions in text and linking them to a knowledge base \cite{xiao2023instructed, li-etal-2020-efficient, Hulst:2020:REL} \\
        & \textbf{Formal translation} & Convert natural language text into formal representations (e.g., SQL, logical forms) \cite{fei-etal-2024-mtls, DBLP:conf/naacl/ZhangGLCGGKWMG24, ganguly2024proof, ryu2025divide} \\
        & \textbf{Recommendation} & Suggest relevant items (e.g., products, documents, movies) \cite{DBLP:conf/sigir/LianHC0WWJFLWC23, DBLP:conf/sigir/ZhaoWXSFC24} \\
        & \textbf{Action execution} & Call APIs based on pre-generated requests \cite{DBLP:conf/nips/SchickDDRLHZCS23, yuan2024craft} \\
        & \textbf{Verification} & Validate model outputs for consistency, logic, and constraint compliance \cite{ganguly2024proof, DBLP:conf/naacl/ZhangGLCGGKWMG24, Olausson_2023}\\
        & \textbf{Aggregation} & When multiple answers available, select/aggregate into one \cite{DBLP:conf/icml/ZhugeWKFKS24, DBLP:journals/corr/abs-2305-17066, Olausson_2023, chu-etal-2024-beamaggr} \\
        & \textbf{GIO creation} & Create a refined, synthesized output that enhances final response for better user interaction; i.e., Generated Information Object (GIO)~\cite{DBLP:journals/sigir/Culpepper0S18}\\
        \cmidrule{2-3}
        & \multicolumn{2}{l}{\cellcolor{gray!15} \textbf{\textit{Complex}}} \\ \cmidrule{2-3}
        & \textbf{RAG} & Combine retrieval and generation to produce factually grounded responses \cite{DBLP:conf/nips/LewisPPPKGKLYR020} \\
        & \textbf{IRCoT} & Integrate retrieval with chain-of-thought for improved information synthesis \cite{DBLP:conf/acl/TrivediBKS23} \\
        & \textbf{Self-RAG} &  Generate text interleaved with reflection tokens to critique and guide the generation process in real-time \cite{DBLP:conf/iclr/AsaiWWSH24} \\
        & \textbf{LINC} & Integrate translation to formal representation and verification for complex logical queries \cite{Olausson_2023} \\
        & \textbf{CoT-decoding} & Elicit intrinsic CoT reasoning in LLMs by exploring top‑k alternative tokens during decoding \cite{DBLP:conf/nips/0002Z24} \\
        \midrule
        \multirow{6}{*}{\textbf{Executors}} 
        & \multicolumn{2}{l}{\cellcolor{gray!15} \textbf{\textit{Agents}}} \\ \cmidrule{2-3}
        & \textbf{Foundation Models and LLMs} & Specific LLMs such as GPT~\cite{brown2020language, OpenAIO1SystemCard2023, OpenAIO3MiniSystemCard2023}, Llama~\cite{DBLP:journals/corr/abs-2302-13971}, T5~\cite{DBLP:journals/jmlr/ChungHLZTFL00BW24},  and multi-modal models \cite{DBLP:conf/iclr/DarcetOMB24, DBLP:conf/icml/RameshPGGVRCS21, DBLP:conf/icml/RadfordKHRGASAM21, DBLP:journals/corr/abs-2312-11805} at different scales and \textit{reasoning} capabilities, or trained with various techniques \\
        \cmidrule{2-3}
        & \multicolumn{2}{l}{\cellcolor{gray!15} \textbf{\textit{Tools}}} \\ \cmidrule{2-3}
        & \textbf{Retrieval models} & Retrieving pieces of information (document text, KG triples, images, etc.) \cite{DBLP:journals/ftir/RobertsonZ09, DBLP:conf/emnlp/KarpukhinOMLWEC20,DBLP:conf/sigir/LiAH24} \\
        & \textbf{Symbolic solvers} & Tools that verify logical statements or prove the validity of expressions using formal reasoning \cite{DBLP:conf/tacas/MouraB08, prover9-mace4} \\
        & \textbf{Other APIs} & External APIs used for various tasks beyond retrieval and reasoning \cite{yuan2024craft} \\
        \midrule
        \multirow{3}{*}{\textbf{Resources}} 
        & \textbf{Document collections} & Internal/external collections of documents, reports, articles, and web pages, possibly in different modalities \\
        & \textbf{Semi-structured resources} & Knowledge graphs with semi-structured sources of factual data (e.g., Wikidata) and databases with predefined schema (e.g., relational SQL databases) \cite{DBLP:conf/semweb/AuerBKLCI07}\\
        \bottomrule
    \end{tabularx}

\end{table*}
}

\subsubsection{Executor modules}
As their name implies, the executor type of modules ($\mathcal{E}$) can perform tasks within our framework.
They serve as the operative entities that process the inputs of a task and decide how to produce outputs that meet the task requirements.
Generally, the combination of tasks and executors determine most of the computational costs of a pipeline.
We propose two categories of executors: \emph{agents} and \emph{tools}.

\textbf{Agents} are general executors capable of performing a large variety of tasks.
Unlike tools, agents are not limited to a few well-defined tasks but instead provide flexibility in application, can reason across multiple inputs and may be capable of utilizing tools when performing tasks.
The archetype of this category are \emph{foundation models} (especially \emph{LLMs}) acting as generalist problem solvers. %

\textbf{Tools} are specialized executors designed to perform specific well-defined tasks with a clear operational scope.
Thus, tools are characterized by their limited applicability due to their specialization, but this often also results in a great efficiency-effectiveness trade-off.
Examples include \emph{classic retrieval models}, \emph{query expansion techniques}, \emph{symbolic solvers}, and various \emph{APIs}.
Tools can constitute efficient pipelines for scenarios that are not too complex.

\subsubsection{Resource modules}
Lastly, resource modules (\(\mathcal{R}\)) define the data sources available to executors.
Generally, resource modules indicate what information is used by a pipeline.
The choice of resources can affect the computational costs incurred, as larger resources may require more time to process, but also other types of costs if a resource is not freely available.
Instead of non-overlapping categories, we categorize by the following properties:

\textbf{Unstructured vs.\ (semi-)structured.}
Whether a resource is structured or not is important to what kind of operations can be performed to it. Structured resources (e.g., \emph{relational databases}, \emph{knowledge graphs}, and \emph{taxonomies}) follow predefined schemata, enabling efficient retrieval, precise querying, and logical inference, which benefit symbolic reasoning and fact verification. In contrast, unstructured resources (e.g., \emph{raw text corpora}, \emph{web documents}, \emph{conversational logs}) lack this structure, making such operations more challenging. However, they are often more abundant and may be the only available resource for certain tasks.

\textbf{Modality.}
The modality of a resource is also important for what tasks and user queries it is useful for.
Since a user can request a specific modality in a response, or task modules may require one or several modalities, e.g., \emph{text}, \emph{images}, \emph{sound}, \emph{video}, etc.

\textbf{Availability: Public, private, or proprietary.}
Finally, the availability of a resource should also be considered.
Ideally a resource is publicly available with no further costs, e.g., \emph{Wikipedia} and \emph{DBPedia}~\cite{DBLP:conf/semweb/AuerBKLCI07}.
However, some resources are private and can only be used for a single user or group of users, e.g., the \emph{user's conversation log} or a \emph{personal} or \emph{corporate document collection}.
Finally, proprietary resources could incur an additional monetary cost if accessed; a further consideration when deciding on a pipeline.

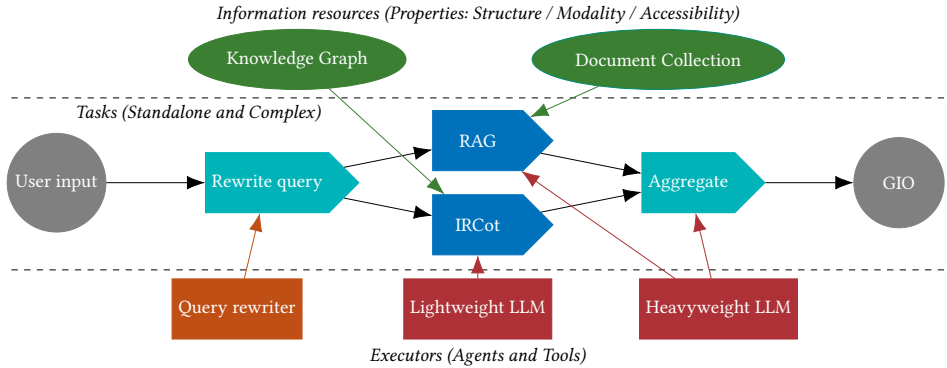
\begin{figure*}[t]
\vspace{-0.75\baselineskip}
\centering
\scalebox{0.8}{
\begin{tikzpicture}
    \node[fill=Bittersweet, text=white, rectangle, draw=Bittersweet, minimum height=1cm, minimum width=1.5cm, signal from=west] at (3,0.1) (rewriter) {Query rewriter};
    \node[fill=Maroon, text=white, rectangle, draw=Maroon, minimum height=1cm, minimum width=1.5cm, signal from=west] at (7,0.1) (llmlight) {Lightweight LLM};
    \node[fill=Maroon, text=white, rectangle, draw=Maroon, minimum height=1cm, minimum width=1.5cm, signal from=west] at (11,0.1) (llmheavy) {Heavyweight LLM};
    \node[fill=gray, text=white, circle, draw=gray, minimum height=1cm, minimum width=1.5cm] at (0,2.2) (input) {User input};
    \node[fill=BlueGreen, text=white, signal, draw=BlueGreen, minimum height=1cm, minimum width=2cm] at (3.5,2.2) (rewrite) {Rewrite query};
    \node[fill=RoyalBlue, text=white, signal, draw=RoyalBlue, minimum height=1cm, minimum width=2cm] at (7,2.9) (rag) {RAG};
    \node[fill=RoyalBlue, text=white, signal, draw=RoyalBlue, minimum height=1cm, minimum width=2cm] at (7,1.5) (ircot) {IRCot};
    \node[fill=BlueGreen, text=white, signal, draw=BlueGreen,, minimum height=1cm, minimum width=2cm] at (10.5,2.2) (aggregate) {Aggregate};
    \node[fill=gray, text=white, circle, draw=gray, minimum height=1cm, minimum width=1.5cm] at (14,2.2) (output) {GIO};
    \node[fill=OliveGreen, text=white, ellipse, draw=PineGreen, minimum height=1cm, minimum width=1.5cm] at (10,4.25) (kgbase) {Document Collection};
    \node[fill=OliveGreen, text=white, ellipse, draw=OliveGreen, minimum height=1cm, minimum width=1.5cm] at (4,4.25) (docs) {Knowledge Graph};
    \draw[dashed] (-0.75,3.61) -- (14.75,3.61);
    \draw[dashed] (-0.75,0.75) -- (14.75,0.75);
    \draw[-{Latex[length=3mm]}] (input) -- (rewrite);
    \draw[-{Latex[length=3mm]}] (rewrite) -- (rag);
    \draw[-{Latex[length=3mm]}] (rewrite) -- (ircot);
    \draw[-{Latex[length=3mm]}] (rag) -- (aggregate);
    \draw[-{Latex[length=3mm]}] (ircot) -- (aggregate);
    \draw[-{Latex[length=3mm]}] (aggregate) -- (output);
    \draw[Bittersweet,-{Latex[length=3mm]}] (rewriter) -- (rewrite);
    \draw[Maroon,-{Latex[length=3mm]}] (llmlight) -- (ircot);
    \draw[Maroon,-{Latex[length=3mm]}] (llmheavy) -- (rag);
    \draw[Maroon,-{Latex[length=3mm]}] (llmheavy) -- (aggregate);
    \draw[OliveGreen,-{Latex[length=3mm]}] (kgbase) -- (rag);
    \draw[OliveGreen,-{Latex[length=3mm]}] (docs) -- (ircot);
    \draw (7,5) node {\emph{Information resources (Properties: Structure / Modality / Accessibility)}};
    \draw (7,-0.7) node {\emph{Executors (Agents and Tools)}};
    \draw (2.35,3.35) node {\emph{Tasks (Standalone and Complex)}};
\end{tikzpicture}
}
\vspace{-0.1\baselineskip}
\caption{
Example of a graph that represents a system design for a possible pipeline in our framework.
Nodes represent modules and are grouped according to their type: \emph{Tasks},  \emph{Executors} and \emph{Resources}, in addition to nodes for the \emph{user's input} and the \emph{output's presentation} (\emph{GIO}~\citep{DBLP:journals/sigir/Culpepper0S18}).
Edges between tasks represent their order of execution and dependencies, edges from executors indicate which are used to execute tasks,
edges from resources indicate which are used for tasks.
}
\end{figure*}

\vspace{-0.5mm}
\subsection{Dynamically creating system pipelines}
\label{sec:framework:pipeline}

The final part of our framework is a methodology that chooses and connects  modules to construct a pipeline.
We formulate a pipeline as a graph where nodes are modules and edges represent the interactions between them.
Uniquely, as per our perspective in Section~\ref{sec:perspective}, we construct a new pipeline for every incoming query on-the-fly, such that our system design is dynamic and adaptive.
We propose that this construction phase is best approached as a planning or multi-step decision process.
Accordingly, we posit that planning and reinforcement learning methods appear to be the best solutions for optimizing the construction phase.
As their design matches the task of learning the best sequence of tasks, assignment of executors, and allocation of resources to optimize an objective trade-off.

\subsubsection{Graph-based representation of pipelines}

Taking inspiration from \citet{DBLP:conf/icml/ZhugeWKFKS24}, we represent each pipeline as a directed acyclic graph (DAG). Let \(\mathcal{G}\) denote the set of all feasible pipeline graphs. Each candidate pipeline \(G\! \in \mathcal{G}\) is a subgraph of the overarching module graph:
\begin{equation}
  G = (V, E), \qquad \text{ where } V \subseteq \mathcal{T} \cup \mathcal{E} \cup \mathcal{R},
\end{equation}
and the directed edges \(e \in E\) capture:
\begin{enumerate*}[label=(\alph*)]
    \item the sequence and flow of information among task modules, and
    \item the assignment/allocation relationships between tasks, executors, and resources.
\end{enumerate*}

Graph configuration appears well-suited for our framing of pipeline construction as a planning problem.
Since, in this framing, constructing a pipeline can amount to sequentially adding nodes and edges until a complete and valid graph is obtained. 

\subsubsection{Query context features}
Our aim is to re-construct and adapt the system's pipeline for each incoming user query. This requires contextual features that are informative about what designs suit a query.
Following contextual bandit terminology, we use a \emph{context vector} $\mathbf{x}_t$ to represent these features for the query of timestep $t$.
It is crucial when defining $\mathbf{x}_t$ to consider what aspects of the query the system should adapt to.
We categorize potential choices for features into three groups:
\begin{enumerate}[leftmargin=*]
    \item \textit{Representations of the user query}:
    Features regarding the user's input query (text) encoding semantic and syntactic aspects or more specific characteristics such as linguistic complexity, indicators of sentiment or emotion, and overall thematic or topical information.
    These should give an indication of query complexity which seems vital for estimating what effectiveness-efficiency trade-off can be made by a pipeline.
    
    \item \textit{User profile features}: This category includes data that characterize the user, such as individual preferences or relevant historical patterns \cite{joko2024doing} or personal data.
    \item \textit{User profile features}: This category includes data that characterize the user, such as individual preferences or relevant historical patterns or personal data \cite{joko2024doing}.
    These features can reduce ambiguity by contextualizing queries, e.g., the typical complexity of a user's questions, or resources they often need \cite{DBLP:conf/sigir/VemuriAMSSSLWPK23}.
    
    \item \textit{Implicit behavior signals}:
    User behavior and interaction patterns often provide a lot of implicit information about the user's current session~\cite{DBLP:conf/uai/HorvitzJH99}.
    For instance, preferences are rarely directly stated but can often be inferred from user interactions.
    Potentially, future systems may even use facial expressions or brain signals to enhance query understanding and provide further information about users' needs~\cite{Ye2025-ps, DBLP:journals/corr/abs-2412-06695, DBLP:conf/sigir/JiSHSS23, Ji_2024, 10.1145/3664647.3681658}.
\end{enumerate}
Together, these features should be indicative of what tasks and resources are needed to respond to the query, efficiently or effectively. %

\subsubsection{The objective of pipeline construction}
\label{sec:framework:pipeline:objective}
The pipeline construction aims to identify a configuration $G \in \mathcal{G}$ that optimally balances efficiency and effectiveness. Formally, we define a composite objective function  $J(G) \in \mathbb{R}$ as a weighted combination of multiple reward and cost signals. 
For the optimization to work, an increase in the objective should indicate a better model, i.e., $J(G) > J(G')$ indicates $G$ is preferred over $G'\!\!$.
We identify four primary sources of feedback on which objectives can be build:
\begin{enumerate}[leftmargin=*]
  \item \textit{User behavior and implicit feedback:}
  User interactions serve as implicit optimization signals.
  Additionally, explicit user annotations, i.e., direct indications of satisfaction with a system response, can give further guidance for improvements.
  \item \textit{Supervised examples:}
  Examples with ground-truth labels, such as manually annotated query resolution pairs or curated gold-standard responses, can be used as target outputs.
  A corresponding objective can measure similarity with the ground-truth as a measure of system effectiveness.
  \item \textit{Self-supervised objectives:}
  Techniques for self-supervision aim to optimize models for goals that do not require labels, but still correlate with effectiveness.
  In our example, a very powerful but inefficient pipeline could provide an example of effectiveness that can be used as feedback to optimize a more efficient pipeline, similar to model distillation \cite{DBLP:conf/ictir/KhandelYVRP24}.
  Furthermore, certain models (e.g., most foundation models) can provide confidence scores about their predicted response.
  Such signals can serve as guidance on the efficiency-effectiveness trade-off of a system \cite{DBLP:journals/nature/FarquharKKG24}.
  For example, low confidence could be an indication that less efficient module additions are required for a query.
  \item \textit{Cost considerations:}
  Efficiency constitutes an important dimension of the our framework's objective, and thus, requires a chosen measure of cost.
  Potential costs to consider include:
  \begin{enumerate*}[label=(\alph*)]
  \item \emph{Response time} or latency of the pipeline, i.e., how many milliseconds do users have to wait for a response to their query;
  \item \emph{Computational costs} of the system, measured in resources like memory, CPU/GPU/TPU FLOPS and network usage.
  \item \emph{Environmental impact} of the pipeline, i.e., its carbon footprint, fresh water usage, etc. \cite{DBLP:conf/sigir/ScellsZZ22}.
  \item \emph{Financial costs} stemming from API calls, hardware usage or proprietary resource access.
  \end{enumerate*}
  These costs are interrelated, and while ideally they would be measured directly, we may have to depend on estimates only.
\end{enumerate}
In addition to average costs, minimizing the variance of these cost measures can enhance robust pipeline behavior in practice. 
We note that feedback signals are rarely perfect; one should consider problems of (selection) bias stemming from feedback sources and transformations from feedback to objectives \cite{DBLP:conf/sigir/GuptaH0VO23, DBLP:conf/sigir/OosterhuisR20}.
\subsubsection{Methodologies for pipeline construction}

A core feature of our proposed framework is the automatic pipeline construction for each individual user query.
This requires a methodology for two core tasks: constructing pipeline graphs on-the-fly based on contextual features, and learning what pipelines work best for which queries.
Thereby, the framework achieves adaptability through selecting and assembling tasks, executors, and resources tailored to individual queries.
Given the combinatorial nature of possible pipelines, we propose that the best suited methodologies come from classical planning and reinforcement learning (RL):
\begin{enumerate}[leftmargin=*]
\item \textit{Classical planning algorithms:}
In scenarios where the effectiveness and efficiency of (partial) pipelines can be estimated accurately, pipeline construction can be approached as a search over possible graph configurations. Adding a module or edge can be seen as traversing a tree of possible graphs. Heuristic and graph-based search methods such as \emph{A*}, \emph{beam search}, and \emph{Monte Carlo tree search} could be effective to determine the optimal configuration \cite{DBLP:journals/tciaig/BrownePWLCRTPSC12}.

\item \textit{General reinforcement learning:}
In order to apply RL approaches, the pipeline construction process has to be formulated as a Markov Decision Process (MDP).
For instance, \emph{states} represent partially constructed pipelines and the query context features; \emph{actions} are the addition of a module nodes or edges;
and \emph{rewards} can come from the objective function (see Section~\ref{sec:framework:pipeline:objective}).
General RL methods can then be applied to the MDP, e.g., \emph{REINFORCE}~\cite{DBLP:journals/ml/Williams92}, \emph{Q-learning}~\cite{DBLP:journals/ml/WatkinsD92} or \emph{actor-critic}~\cite{DBLP:conf/nips/KondaT99} approaches.

\item \textit{Contextual bandit algorithms:}
In scenarios where the number of possible graphs is limited, contextual bandit algorithms become relevant.
This subcategory of RL algorithms works for cases where the number of actions is discrete and rewards are immediate and only depend on single actions.
Whilst these methods only apply in a subspace of all possible settings for RL, their specialization enables them to learn from far fewer examples than general RL methods, thus allowing for increased adaptivity.
If the number of modules is limited, one can pre-compute all possible graphs and use them as the set of actions for CMAB algorithm.
\end{enumerate}

\vspace{-1mm}
\subsection{Limitations and potential extensions}

Before diving into a specific instantiation of our framework, using CMAB, we consider future extensions that would address limitations of our proposal.

To start, reliance on RL methods that respond to contextual features comes with two possible limitations.
First, the data-efficiency of RL methods could be too low to be responsive, i.e., if the RL method is too slow to learn a new graph construction strategy, it may already have been outdated before it is learned.
This problem is most likely when the number of possible graphs is enormous, since RL methods have difficulty with large action spaces.
Second, the contextual features may not be sufficiently informative to enable effective adaptivity, i.e., the query context features may not enable the identification of the best graph structure.
Both are general problems with the RL methodology, where promising advancements in the field of RL may help alleviate these limitations~\cite{DBLP:journals/tmlr/Li0LZLY23, DBLP:conf/icml/BauerBBBBCCCDGG23}.

Also, our framework is designed to construct a graph for each query for a given set of modules.
The optimization for underlying modules is taken for granted, and the constructed graphs are only based on features that are available when a queries are issued.
Obvious directions of extension include the joint optimization of graph construction and underlying modules, e.g., by incorporating prompt and model fine-tuning in the learning process; and, enabling graphs that are dynamic during execution.
The latter would enable graphs that adapt to intermediate outcomes, e.g., a first task could be to retrieve a result, and the subsequent tasks are chosen based on that first result.
Whilst this would greatly increase the complexity of the graph construction process, it could be a logical continuation for development of our framework.

\section{An Instantiation of our Framework with a Contextual Bandit Approach}
\label{sec:Instantiation}

To illustrate how our framework could be applied, we provide an example instantiation for a question-answering (QA) task. %
\vspace{-1mm}
\subsection{Dataset}
\label{sec:instantiation:dataset}
We use the dataset developed by \citet{DBLP:conf/naacl/JeongBCHP24} for building and evaluating RAG for QA systems.
It features a diverse set of questions, each labeled for one of three levels of complexity using Flan-T5-XL.
Label A indicates that the question can be answered directly without retrieval;
label B requires single-step retrieval;
and label C necessitates multiple reasoning and retrieval steps. 
For our study, we randomly select 210 training and 51 test questions, ensuring a balanced distribution across complexity levels for optimization.

\subsection{Graph definition}
\label{sec:instantiation:graph}
Our choice of nodes and possible edges matches the setup proposed by \citet{DBLP:conf/acl/TrivediBKS23} and used by \citet{DBLP:conf/naacl/JeongBCHP24}. 
We define three complex tasks and one standalone task:
\begin{equation}
\mathcal{T} = \{ \textit{NoR}, \textit{OneR}, \textit{IRCoT}, \textit{Aggregate} \}.
\end{equation}

\begin{itemize}[leftmargin=*]
\item \textbf{NoR}:
Answer generation with \textit{\textbf{no} \textbf{r}etrieval} augmentation, to be generated by an LLM agent module.
The NoR task is to answer questions without additional external data, making it most suited for simple non-knowledge-intensive questions that match its encoded parametric knowledge.
\item \textbf{OneR}:
Answer generation with \textit{\textbf{one}-time \textbf{r}etrieval}, to be generated by an LLM agent module with a retriever tool and resource.
During this task, a retriever tool is called \emph{once} to provide seven potentially relevant documents to the executor. %
OneR is suited for questions that require access to external knowledge. 
\item \textbf{IRCoT}:
Answer generation with following the IRCoT procedure, performed by an LLM agent module with a retriever tool and resource.
Retrieved documents are interleaved in several back-and-forth chain-of-thought (CoT) reasoning steps with the LLM. %
As the most complex and time-consuming strategy, IRCoT is most-suited for questions where NoR and OneR are expected to fail, i.e., ones that demand extensive knowledge or a synthesis of multiple pieces of knowledge to produce an accurate answer.
\item \textbf{Aggregate:}
Aggregating the answers from multiple tasks into a single coherent answer, to be performed by an LLM or aggregator tool, i.e., a simple rule-based function.
\end{itemize}
The \textit{NoR}, \textit{OneR}, \textit{IRCoT} tasks can only be executed in parallel, the \textit{Aggregate} task can solely process the output of the other tasks.

For executor modules, we include Flan-T5-XL as an \emph{LLM agent}, and BM25 as a \textit{retrieval tool module}~\cite{DBLP:conf/trec/RobertsonWJHG94},
and an \textit{aggregator tool} that applies a majority voting function to select the most frequent answers from connected tasks.
We avoid LLM-based aggregators as they may introduce errors or hallucinations.\footnote{See \citep{DBLP:conf/icml/ZhugeWKFKS24, DBLP:journals/corr/abs-2305-17066} for other possible aggregation modules and their purposes.}

Finally, the resource module includes two corpora: \emph{Wikipedia}~\cite{DBLP:conf/emnlp/KarpukhinOMLWEC20} and a \textit{passage corpus} constructed for multihop questions~\citep{DBLP:conf/acl/TrivediBKS23} .

\subsection{Optimization with CMAB}
\label{sec:instantiation:method}
Due to the limited number of possible valid graphs in our setup (seven in total), we choose to apply the LinUCB contextual multi-armed bandit (CMAB) algorithm~\citep{li2010contextual}.
Our contextual features are the complexity labels provided in the dataset that are indicative of what kind of pipeline is required to answer each question.
The reward function is a linear combination between \emph{correctness} ($P_t$), evaluated by F1-score, and costs from compute time ($T_t$):
\begin{equation}\label{eq:reward}
r_t = \beta \cdot P_t - (1-\beta) \cdot T_t,
\end{equation}
we set $\beta=0.5$ and $T_t = S_t \big(\frac{\mathds{1}[1 < S_t \leq 10]}{10000} +  \frac{\mathds{1}[S_t > 10]}{50}\big)$ where $S_t$ is execution time in seconds.
Our choice of $T_t$ represents that we do not care for times below a single second, and really want to avoid times over ten seconds.
We start with a pre-computing phase where every valid graph is generated and gathered in a set, this results in seven valid graphs that serve as the arms of the CMAB.
Subsequently, the CMAB process is started:
at each timestep $t$ a random question is sampled from the training set,
based on its context features CMAB chooses an arm which determines the system pipeline used to answer the query.
The reward for the resulting answer is computed and CMAB updates its parameters accordingly, and its parameters are evaluated on the test-set.
This process is repeated for 3,500 timesteps, during which CMAB aims to learn which system pipelines best match which question complexities.

\begin{table}[t]
\centering
\caption{Performance of each individual agent on the training set (by F1-score and average time in seconds).}
\label{tab:train_eval}
\setlength{\tabcolsep}{3.5pt} 
\begin{tabular}{l cccccc}
\toprule
 & \multicolumn{2}{c}{\textbf{NoR}} & \multicolumn{2}{c}{\textbf{OneR}} & \multicolumn{2}{c}{\textbf{IRCoT}} \\
\cmidrule(lr){2-3} \cmidrule(lr){4-5} \cmidrule(lr){6-7}
 & {F1} & {Time (s)} & {F1}  & {Time (s)}  & {F1}  & {Time (s)} \\
\midrule
\textbf{Context A} & 0.914 & 0.66 & 0.677 & 6.46 & 0.730 & 189.78  \\
\textbf{Context B} & 0.061 & 0.66 & 0.518 & 7.34 & 0.580 & 192.30  \\
\textbf{Context C} & 0.066 & 0.67 & 0.146 & 6.41 & 0.458 & 184.85  \\
\textbf{Overall} & 0.347 & 0.66 & 0.447 & 6.74 & 0.589 & 188.97  \\
\bottomrule
\end{tabular}
\end{table}

\subsection{Experimental results}
\label{sec:results}

Here we present results for the instantiated problem in Section \ref{sec:Instantiation}.
\noindent
To interpret the results of our CMAB experiments effectively later, we first assess the individual performance of each Task defined in $T$ on the training dataset. As shown in Table \ref{tab:train_eval}, performance varies by complexity of inputs: NoR excels in simple cases, while IRCoT dominates at higher complexities (B and C). At level B, IRCoT slightly outperforms OneR (.580 vs. .518) but incurs significantly higher latency (192.30 vs. 6.41s) due to iterative retrieval and reasoning. This trade-off highlights the practical cost of improved accuracy, as increased response time translates to the waiting time for the user to get a potentially correct response from the system.

\begin{table}[t]
\centering
\caption{Evaluation of AQA (NT: Time-agnostic reward, T: Time-based reward) and GPTSwarm \citep{DBLP:conf/icml/ZhugeWKFKS24} on the test set by F1-score and time (log-transformed, in ms). Both have been trained on the training set. For GPTSwarm, the final optimized graph configuration is used for evaluation. 
}
\label{tab:test_eval}
\setlength{\tabcolsep}{3.5pt} 
\begin{tabular}{l cc cc cc}
\toprule
 & \multicolumn{2}{c}{\textbf{AQA (NT)}} & \multicolumn{2}{c}{\textbf{AQA (T)}} & \multicolumn{2}{c}{\textbf{GPTSwarm}} \\
\cmidrule(lr){2-3} \cmidrule(lr){4-5} \cmidrule(lr){6-7}
 & {F1} & {Time} & {F1} & {Time} & {F1}  & {Time} \\
\midrule
\textbf{Context A} & 1.0\phantom{00} & \phantom{0}6.18 & 1.0\phantom{00} & \phantom{0}6.18 & 0.862 & 12.78 \\
\textbf{Context B} & 0.568 & 12.04 & 0.539 & \phantom{0}8.73 & 0.327 & 12.79 \\
\textbf{Context C} & 0.523 & 11.75 & 0.523 & 11.75 & 0.317 & 12.76 \\
\textbf{Overall} & 0.697 & \phantom{0}9.99 & 0.687 & \phantom{0}8.89 & 0.502 & 12.78 \\
\bottomrule
\end{tabular}
\end{table}

\begin{figure*}[t]
\centering
\raisebox{-.5\height}{
\includegraphics[width=0.75\textwidth]{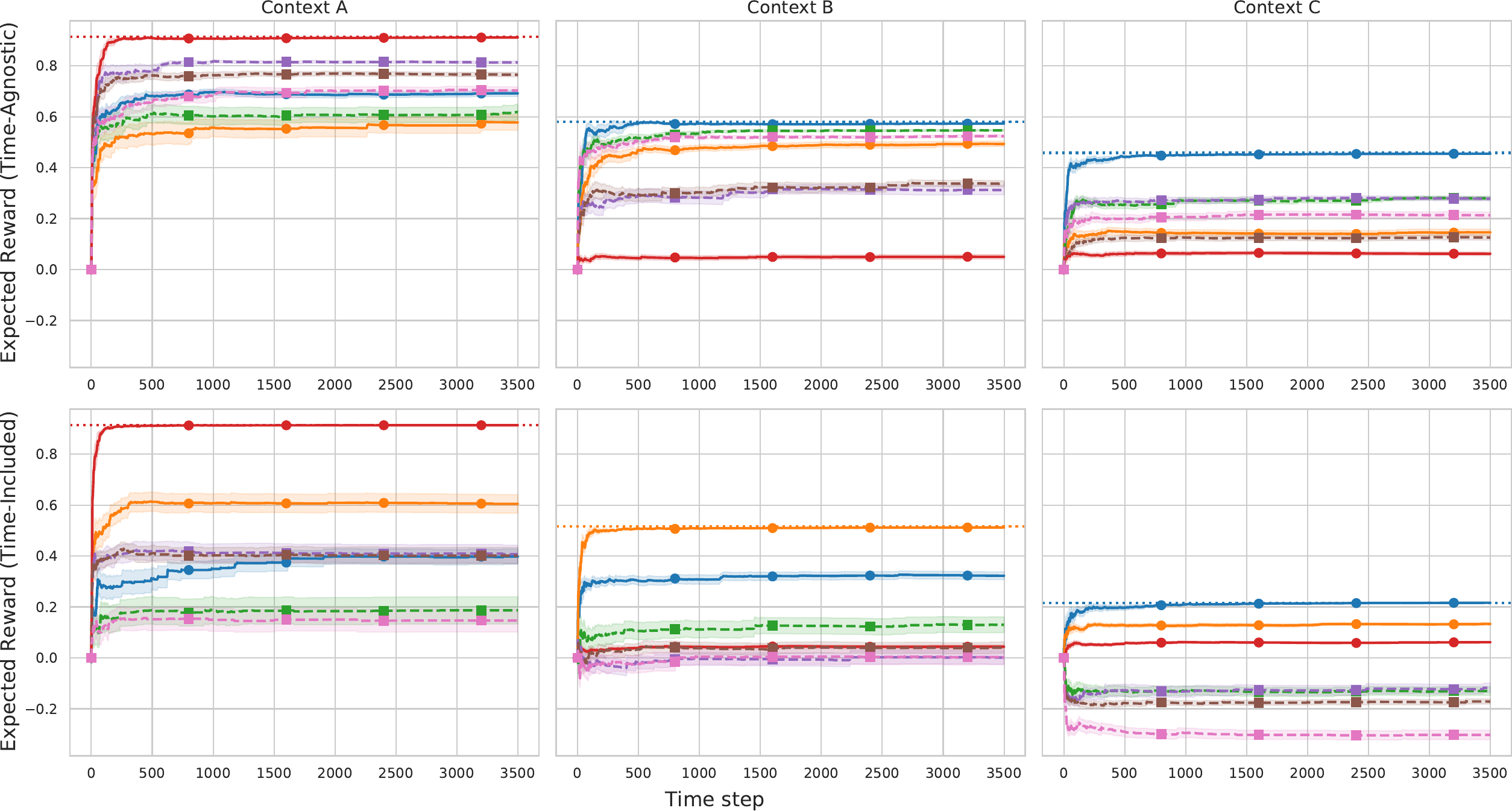}
}\hspace{-2pt}%
\raisebox{-.5\height}{
\includegraphics[width=0.15\textwidth, trim=1.4em 0 1.65em 0, clip]{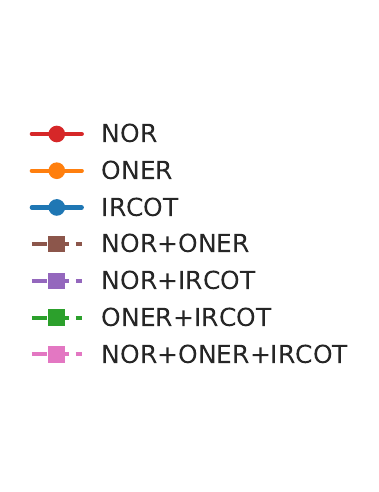}
}
\caption{LinUCB expected rewards for the collaborative action space, dashed line depicts real reward for the optimal action.}
\label{fig3}
\end{figure*}

\subsubsection{Can CMAB learn to map query complexity to optimal pipeline structures?}
We aim to verify whether CMAB can learn to adaptively map different levels of input complexities to the optimal graph configurations.

\begin{enumerate}[label=(\alph*),leftmargin=*]
\item \textbf{Time-agnostic reward.}
In Figure \ref{fig3}, the top row depict the expected rewards for LinUCB, without considering time in the reward. These results reveal optimal actions identified for each complexity label; NoR for context A, and IRCoT for contexts B and C, based on F1-scores. The dotted lines depict the real reward calculated for the best possible action per context.

\item \textbf{Time-based reward.} 
The bottom row of Figure \ref{fig3} depicts results for training while including time cost in the reward calculation. Considering both performance and the time cost, the real reward for best actions per label are shown in the plots as dotted red lines. Including time in the reward calculation, we observe the model preferentially selects pipelines that balance efficacy with reasonable execution times. In context B, it prioritizes the configuration with only OneR over configurations with IRCoT or a combination of better-performing agents (i.e., OneR+IRCoT).
\end{enumerate}
We conclude that our approach successfully constructs system pipelines that are adapted to the characteristics of incoming questions and optimized for a given efficiency-effectiveness trade-off.

\subsubsection{Can adaptive orchestration achieve a superior trade-off over static optimization?}
To demonstrate the importance of adaptivity in designing an orchestration framework, here, we aim to assess how a non-adaptive orchestration optimization affects the performance. To this aim, we train our \emph{adaptive question-answering model (AQA)} and GPTSwarm \cite{DBLP:conf/icml/ZhugeWKFKS24} and assess them on our test set. 
Both models are given the set of modules described in Section~\ref{sec:instantiation:graph}. Since GPTSwarm does not accommodate context or time cost in their proposed framework, in our experiment with this framework we feed each question to the model and the optimization gradually converges to the \textit{most optimal} graph configuration by optimizing towards better performance (F1-score).\footnote{Similar to what is done in their MMLU experiment \cite{DBLP:conf/icml/ZhugeWKFKS24}.} 

Prior to optimization with REINFORCE, the graph maintains a uniform edge probability distribution. After 200 training epochs, optimization converges on a structure where only OneR and IRCoT nodes retain edges, while OneR's connection to the final decision node is pruned out (following the thresholding strategy of \cite{DBLP:conf/icml/ZhugeWKFKS24}, edges with probabilities below .5 are removed). This shows, as expected, that this static framework fails to adapt to different complexity levels by routing for a good enough answering strategy that is also time-efficient, as it only favors higher accumulative F1 scores across time.

The evaluation results on the test set are shown in Table~\ref{tab:test_eval}. The first two columns show the CMAB model (LinUCB) with and without time considered in the reward implementation respectively and the last column is the final optimized graph using REINFORCE. 
Using a fixed orchestration by GPTSwarm, we observe a fall in performance for different complexity levels, as it is not possible to adapt the strategy based on the characteristics of the question, which leads to a lower overall performance compared to AQA (NT and T). Also, as the fixed configuration in GPTSwarm is also more sophisticated (i.e., both NoR and IRCoT have edges to the final decision node), the time cost is constantly higher compared to AQA.

\vspace{-2mm}
\section{Summary and Future Research Directions}
\label{sec:conclusion}
In this paper, we recognize that the landscape of generative information access systems is marked by the proliferation of heterogeneous modules and tools -- each designed to serve similar purposes, yet emerging as isolated implementations. Observing this fragmented state, we raised and addressed critical questions regarding the design of a comprehensive framework that allows for the inclusion of these heterogeneous modules.
We posit that in order to move forward, module orchestration is not enough; the field needs to develop methods for \emph{adaptive} module orchestration. The ultimate goal should be that the optimal answering strategy can be selected for each user input, so that the most relevant answer is inferred at the lowest possible cost.
In this regard, we proposed a future-ready framework that defines the potential modules (i.e. building blocks) of a GenIA system as nodes of a graph with dynamic edges, enabling adaptive orchestration of pipelines in real-time. 

Our perspective bridges existing gaps by introducing a formalized approach to adaptive orchestration, ensuring that chosen answering strategies are context-aware and resource-efficient.
To realize such a comprehensive self-organizing adaptive system built on many subcomponents, we have outlined how such a framework and its optimization can be modeled.
To indicate the practical usage of our framework, we provided a proof-of-concept instantiation of this framework for a query-answering task.
Our experimental evaluation of this instantiation indicated that our approach can successfully construct system pipelines that are adapted to the characteristics of incoming questions and optimized for a given efficiency-effectiveness trade-off.

Looking ahead, we like to pose some open questions for the community and some future research directions.
\begin{enumerate*}[label=(\roman*)]
    \item As the system scales by integrating more modules, how should optimization methods adapt to ensure efficiency and robustness?
    \item What internal signals from system subcomponents (e.g., uncertainty or likelihood signals) can enhance decision-making at each step? How should these signals be leveraged effectively?
    \item How to handle orchestration complexity that arises from more granular node definitions and task decompositions?
    \item How can optimization models be designed to directly learn dependencies between context features and graph configurations, ensuring adaptive orchestration without relying on intermediate classification/clustering of the queries?
    \item How and what additional contextual signals should inform decision-making?
    \item How can we account for other levels of optimization (e.g., enhancing prompts' quality or fine-tuning at individual module level)?
    \item As the number of optimization goals increases, how can we ensure effective convergence without compromising performance?
\end{enumerate*}
These questions opens up interesting research directions and new challenges, that may bring us closer to achieving self-organizing conversational assistants that automatically adapt to different environments while maintaining robustness and flexibility.

\medskip
\noindent\textbf{\textit{Resources.}}
Resources needed to reproduce the results of this paper are publicly available at \url{https://github.com/informagi/AQA}.

\begin{acks}
We thank Nik Vaessen for his intellectual contributions to help shape our ideas.
This research is supported by the Dutch Research Council (NWO) under project numbers NWA.1389.20.183, VI.Veni.222.269, 024.004\-.022, and KICH3.LTP.20.\-006, and the EU’s Horizon Europe program under grant No.\ 101070212 and No.\ 101070014
(OpenWebSearch.EU, \url{https://doi.org/10.3030/101070014}).
All content represents the opinion of the authors, which is not necessarily shared or endorsed by their respective employers and/or sponsors.
\end{acks}

\balance
% \bibliographystyle{ACM-Reference-Format}
% \bibliography{references}

%%% -*-BibTeX-*-
%%% Do NOT edit. File created by BibTeX with style
%%% ACM-Reference-Format-Journals [18-Jan-2012].

\end{document}